\documentclass[%
 apl,
 amsmath,amssymb,
preprint,%
]{revtex4-1}

\usepackage{graphicx}% Include figure files
\usepackage{dcolumn}% Align table columns on decimal point
\usepackage{bm}% bold math
\usepackage[utf8]{inputenc}
\usepackage[T1]{fontenc}
\usepackage{mathptmx}

\usepackage{color}
\usepackage{sidecap}

\newcommand{\YBCO}{$\mathrm{YBa}_2\mathrm{Cu}_3\mathrm{O}_7$\;}

\begin{document}

\preprint{AIP/123-QED}

\title[High-$T_c$ superconducting detector for highly-sensitive microwave magnetometry]{High-$T_c$ superconducting detector for highly-sensitive microwave magnetometry\\}
% Force line breaks with \\

%%\author{A. Author}
 %%\altaffiliation[Also at ]{Physics Department, XYZ University.}%Lines break automatically or can be forced with \\
%%\author{B. Author}%
%% \email{Second.Author@institution.edu.}
%%\affiliation{ 
%%Authors' institution and/or address%\\This line break forced with \textbackslash\textbackslash
%%}%

%%\author{C. Author}
%% \homepage{http://www.Second.institution.edu/~Charlie.Author.}
%%\affiliation{%
%%Second institution and/or address%\\This line break forced% with \\
%%}%

\author{Fran\c{c}ois Cou\"edo,$^{1}$ Eliana Recoba Pawlowski,$^{2}$ Julien Kermorvant,$^{3}$ Juan Trastoy,$^{2}$ Denis Cr\'{e}t\'{e},$^{2}$ Yves Lema\^{i}tre,$^{2}$ Bruno Marcilhac,$^{2}$ Christian Ulysse,$^{4}$ Cheryl Feuillet-Palma,$^{1}$ Nicolas Bergeal,$^{1}$ }
\author{J\'{e}rome Lesueur,$^{1}$}
\email[Corresponding author : ]{jerome.lesueur@espci.fr}
\affiliation{$^{1}$ Laboratoire de Physique et d'Etude des Mat\'eriaux, CNRS, ESPCI Paris, PSL Research University, UPMC, 75005 Paris, France.}
\affiliation{$^{2}$ Unit\'e Mixte de Physique CNRS, Thales, Universit\'e Paris-Sud, Universit\'e Paris-Saclay, 91 767 Palaiseau Cedex, France.}
\affiliation{$^{3}$ Thales Communication and Security, 92230 Gennevilliers, France}
\affiliation{$^{4}$ Centre de Nanosciences et de Nanotechnologie, CNRS, Universit\'e Paris Saclay, 91120 Palaiseau, France.}

\date{\today}% It is always \today, today,

\begin{abstract}
We have fabricated arrays of High-$T_c$ Superconducting Quantum Interference Devices (SQUIDs) with randomly distributed loop sizes as sensitive detectors for Radio-Frequency (RF) waves. These sub-wavelength size devices known as Superconducting Quantum Interference Filters (SQIFs) detect the magnetic component of the electromagnetic field. 
We used a scalable ion irradiation technique to pattern the circuits and engineer the Josephson junctions needed to make SQUIDs.
Here we report on a 300 SQUIDs series array with loops area ranging from $6$ to $60\ \mu \mathrm{m}^{2}$, folded in a meander line covering a $3.5\ $mm$\times 120\ \mu$m substrate area, made out of a $150\ $nm thick \YBCO (YBCO) film. Operating at a temperature $T=66\ $K in an un-shielded magnetic environment, under low DC bias current ($I=60\ \mu$A) and DC magnetic field ($B=3\ \mu$T), this SQIF can detect a magnetic field of a few pT at a frequency of $1.125\ $GHz, which corresponds to a sensitivity of a few hundreds of fT/$\sqrt{\mathrm{Hz}}$, and shows linear response over 7 decades in RF power. 
This work is a promising approach for the realization of low dissipative sub-wavelength GHz magnetometers.
\end{abstract}

\maketitle

Detecting electromagnetic fields in the micro-wave domain with high precision and resolution is a  pivotal issue for both basic science and applications. For instance in solid state studies, highly sensitive magnetometers are needed to detect electromagnetic fields in quantum information systems\cite{Kolkowitz:2015fx}, to study electron spin dynamics\cite{Hall:2015dl} or to make advanced Nuclear Magnetic Resonance spectroscopy \cite{Kimmich:2004ie}. At the same time, further progress in security systems, wireless and satellite communications or radars requires significant improvement of state of the art Radio Frequency (RF) detectors\cite{Gameiro:2018de}.

While most of the electromagnetic wave detectors are based on a resonant electrical dipole for enhanced sensitivity, the need of sub-wavelength devices is increasing, to miniaturize the detectors and include them in compact and mobile ensembles, or to image electromagnetic fields at small scale\cite{Wahnschaffe:2017jj,Vlaminck:2012by,Thiel:2016es}. Such lumped elements are usually broadband\cite{Kornev:2009fja}, which is of high interest for many applications. One way to fulfill all these requirements (high sensitivity, sub-wavelength size, broadband operation) is to detect the magnetic component of the wave instead of the electric one as usual.

The state-of-the-art magnetometers reach their best sensitivity in a narrow band of frequency, and  typically operate at frequencies lower than 10-100 MHz. Magnetic sensitivity in the range of (sometimes sub-) fT/$\sqrt{\mathrm{Hz}}$ can be achieved\cite{Stark:2017jr,Lee:2006cw,Storm:2017jr} using two technologies : atomic magnetometers\cite{Weis:2005iu} and Superconducting Quantum Interference Devices (SQUIDs)\cite{Clarke:2005tz}. While the former are based on optical transition between magnetic sensitive atomic levels,  the latter rely on quantum interferences in a superconducting loop interrupted by Josephson Junctions (JJ). The high and comparable sensitivities of both systems hold at low frequency and rapidly degrade beyond typically a few MHz. Using Nitrogen Vacancy (NV) centers in diamond, Stark \textit{et al} reported a sensitivity of $1\ \mu \mathrm{T}/\sqrt{\mathrm{Hz}}$ at 1.6 GHz\cite{Stark:2017jr}, while Horsley \textit{et al} reached $1.8\ \mu \mathrm{T}/\sqrt{\mathrm{Hz}}$ up to 26 GHz with a Rb atomic vapor cell\cite{Horsley:2016ce} and $130\ \mathrm{nT}/\sqrt{\mathrm{Hz}}$ at 2.7 GHz with NV centers in diamonds\cite{Andrew:2018fo}.

Limitations for high frequency operation also hold for SQUIDs. Indeed, these devices have a response which is $\Phi_{0}$ periodic in applied magnetic flux ($\Phi_{0}=\frac{\textit{h}}{2e}$ being the flux quantum). An external feedback loop is used to operate them in a limited range of magnetic field corresponding to a single valued response. The dynamics of the feedback electronics limits in practice the SQUIDs bandwidth to 100 MHz\cite{Hilbert:1985ja} in the best cases, unless a special implementation is used to reach up to GHz, but in a severely reduced bandwidth\cite{Muck:2003ia}. To overcome this drawbacks, Superconducting Quantum Interference Filters (SQIFs), were developed  since the pioneer work of Oppenl\"{a}nder \textit{et al}\cite{Oppenlander:2000hz}. A SQIF is an array of SQUIDs with incommensurate loops sizes. All the periodic responses of individual SQUIDs cancel, and the voltage response to an applied external magnetic field is single valued. Thus, the magnetic DC response is highly peaked and symmetric around zero magnetic field, with an extended linear part\cite{Mukhanov:2014ig,Kornev:2017kra,Cybart:2017bx,Muck:2010cd}. As a consequence, there is no need of feedback electronics, and therefore no limitation in frequency for this reason. A SQIF is therefore an absolute magnetometer, with an interesting linear range when magnetically polarized around the maximum slope of the peak, and potentially operating at high frequency.  Moreover, such devices can combine different roles and serve as sub-wavelength antennas and amplifiers for instance \cite{Kornev:2017hva}. Based on the well-established technology of niobium-based Josephson Junctions (JJ) and optimized architectures \cite{Kornev:2017hva,Kornev:2007bm}, SQIFs were successfully operated as RF detectors up to 15 GHz with gain in the 20-25 dB range\cite{Prokopenko:2013to,Prokopenko:2015dx}. High-$T_c$ Superconductors (HTS) have also been used to make SQIFs\cite{Shadrin:2008kka,Cybart:2014et,Mitchell:2016in,Ouanani:2016cr} but the maximum operation frequency reported to date is  about 100-200 MHz\cite{Snigirev:2007ih,Kalabukhov:2008id,Pawlowski:2018jj}. In the present article, we report on an HTS SQIF operating in the GHz frequency range with a sensitivity in the hundreds of $\mathrm{fT}/\sqrt{\mathrm{Hz}}$ range.

\begin{figure}[h]
\includegraphics[scale=0.8]{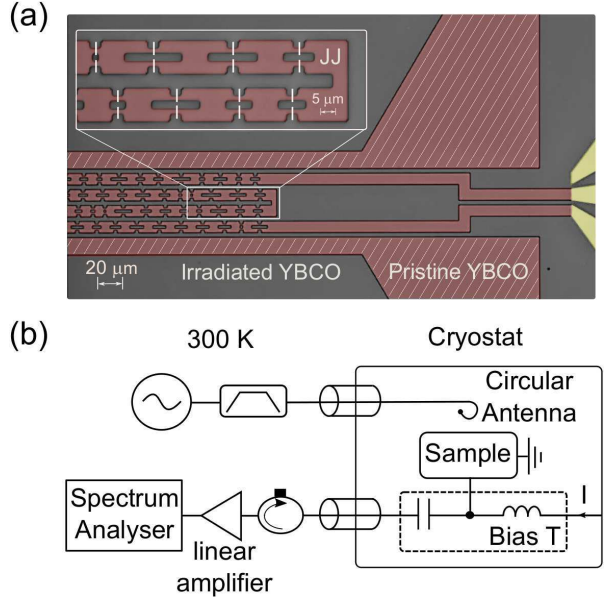}\\
  \caption{
  \textit{(a)} Optical picture of the device. SQUIDs are in series in a meander line to make the SQIF (central part of the picture), which is connected to the CPW (CoPlanar Waveguide) line made of gold covered YBCO (on the right part of the picture). JJ are indicated by white lines in the blow-up part. The regions colored in brown are pristine and superconducting. The ones in grey are insulating. The superconducting line around the SQIF (dashed line texture) is not used in this experiment.
  \textit{(b)} Sketch of the measurement set-up. The RF input signal is sent onto the sample through a circular antenna placed above it. The detected signal is conveyed through an on-chip CPW line followed by a coaxial cable to the room-temperature amplifier and the spectrum analyzer. A bias-T is used to DC bias the sample.
  }
  \label{Figure1}
\end{figure}

We have fabricated an HTS SQIF using the ion irradiation technique in a two-step process (details on the fabrication techniques can be found in our previous papers\cite{Bergeal:2005jna,Bergeal:2006dd,Bergeal:2007jc,Ouanani:2014cu,Ouanani:2016cr,Pawlowski:2018jj}). Starting from a commercial\footnote[1]{Ceraco Gmbh.} 150 nm thick c-axis oriented \YBCO (YBCO) film grown on a sapphire substrate, we first design the superconducting circuit, namely the SQUID rings, their interconnections and the contact pads : a photoresist mask protects the film from a 110 keV oxygen ion irradiation at a dose of $5\times10^{15}\ $ions/cm$^{2}$ to keep it superconducting. The unprotected part becomes insulating. In a second step, an e-beam sensitive resist mask covering the whole film is used, in which trenches of 40 nm  wide and a few microns long have been opened at places where Josephson junctions (JJ) will be created, namely accross the SQUIDs arms. Another ion irradiation at much lower dose ($3\times10^{13}\ $ions/cm$^{2}$) defines the JJ by lowering locally the superconducting $T_{c}$. The SQIF used in this study is made of 300 SQUIDs in series, with loop sizes randomly distributed between 6 and 60 $\mu \mathrm{m}^{2}$ folded in a meander line (see Figure \ref{Figure1} \textit{(a)}). The width of the SQUIDs arms is 2 $\mu$m in the vicinity of the JJs. The total size of the device is $8\ $mm by $120\ \mu$m. This is much smaller than the wavelength $\lambda \sim\ 30\ \mathrm{cm}$ of the RF waves used in this experiment (frequency $\sim 1\ $GHz). The SQIF can therefore be considered as a lumped element at such frequencies.

% Figure 1

The SQIF is mounted on a Printed-Circuit Board (PCB) with a Co-Planar Wave guide (CPW) transmission line. It is then placed in a magnetically un-shielded cryogen-free cryostat, equipped with Helmholtz coils to generate a DC magnetic field, filtered wires to DC bias the device and coaxial cables for RF measurements.
The measurement setup is schematically shown in Figure \ref{Figure1} \textit{(b)}. The input RF signal is generated in a continuous mode at a fixed frequency $f$ and a given source power $P_{RF}$, band-filtered at room temperature (in a 400 MHz bandwidth (around 1.17 GHz)) and then coupled to the sample through a circular antenna (5 mm in diameter) placed at about 1 cm from the SQIF surface, \textit{i.e.} in a near-field condition for $\sim 1\ $GHz frequency wave in vacuum. The RF output signal is isolated from DC by a bias-tee at low temperature, pre-amplified at room temperature (+ 40 dB gain) and measured with a spectrum analyzer in a zero-span mode with a 1-kHz resolution bandwidth. A circulator has been used to prevent the amplifier's noise to radiate back onto the sample.

\begin{figure}[h]
\includegraphics[scale=0.7]{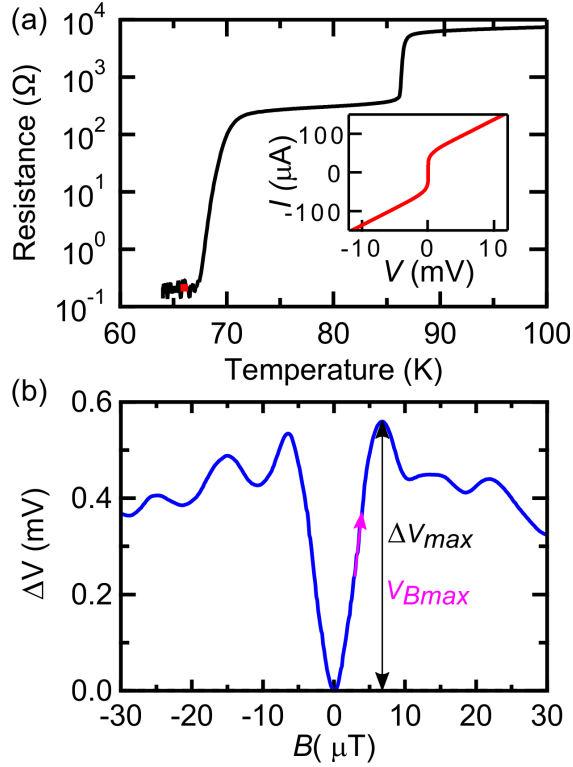}\\
  \caption{
  \textit{(a)} Resistance $R$  of the SQIF as a function of temperature $T$. 
  \textit{(Inset)} : Current $I$ versus voltage $V$ of the device at $T=66\ $ K (red point on the main panel). 
  \textit{(b)} Voltage difference $\Delta V=V-V_{min}$ ($V_{min}=1.37\ \mathrm{mV}$) as a function of the applied DC magnetic field $B$ at $T=66$ K, for a bias current of $I=60\ \mu$A. The maximum voltage swing $\Delta V_{max}=max(\Delta V)$ and the maximum transfer function $V_{Bmax}=max\left(\frac{\partial V}{\partial B}\right)$ are shown by arrows.
  }
  \label{Figure2}
\end{figure}

Figure \ref{Figure2} \textit{(a)} shows the resistance $R$ of the SQIF as a function of temperature $T$. As reported previously \cite{Bergeal:2005jna,Ouanani:2014cu,Ouanani:2016cr}, a Josephson behavior is observed below a coupling temperature $T_J=67$ K. The normal state resistance measured at $T=80$ K is $R_N= 309$ $ \Omega$. The Current-Voltage (IV) characteristic (\textit{inset} Figure \ref{Figure2} \textit{(a)}) measured at $T=66$ K is typical of ion irradiated YBCO JJ\cite{Katz:2000gf,Bergeal:2007jc,Malnou:2012gt,Ouanani:2016cr}. 

% Figure 2

%%%%%%%%%%%

% Figure 3
\begin{figure*}[t]
%
%\begin{minipage}{0.67\textwidth}
%
\includegraphics[scale=0.57]{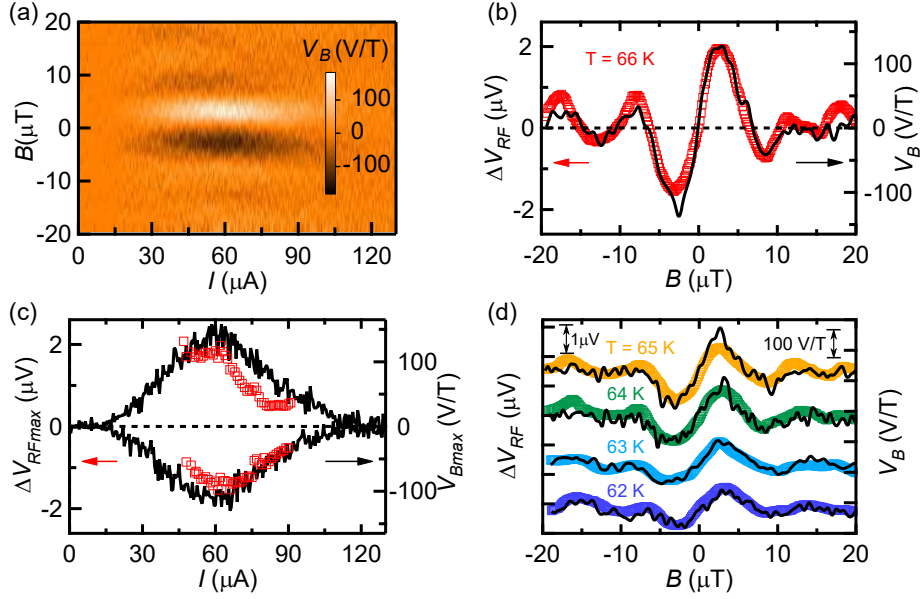}%\\
 %
% \end{minipage}\hfill
% \begin{minipage}{0.285\textwidth}
 %
  \caption{
  \textit{(a)} 2D plot of the transfer function $V_{B}=\frac{\partial V}{\partial B}$ as a function of the bias current $I$ and applied DC magnetic field $B$ at $T_{OPT}=66$ K.
\textit{(b)} Measures at $T_{OPT}$. Red symbols, left axis : RF voltage difference $\Delta V_{RF}=V_{RF}(B)-V_{RF}(B=0)$ as a function of $B$ at $I_{OPT}$ ($P_{RF}=0\ \mathrm{dBm}$). Black line, right axis : Cut of the 2D plot $I_{OPT}=60\ \mu$A : $V_{B}$ as a function of $B$. 
\textit{(c)} Measures at $T_{OPT}$. Red symbols, left axis : $\Delta V_{RF}^{-}=min(\Delta V_{RF}(B<0))$ and $\Delta V_{RF}^{+}=max(\Delta V_{RF}(B>0))$  as a function of $I$ ($P_{RF}=0$ dBm). Black line, right axis : $V_{B}^{-}=min(V_{B}(B<0))$ and $V_{B}^{+}=max(V_{B}(B>0))$ as a function of $I$.
\textit{(d)} Same plot as panel \textit{(b)} for different temperatures below $T_{J}$. For the sake of clarity, curves have been vertically shifted ($1\ \mu$V for $\Delta V_{RF}$ and $100$ V/T for $V_{B}$).
}
  \label{Figure3}
  %
%  \end{minipage}
  %
\end{figure*}

%%%%%%
The device biased above its critical current $I_c$ displays a typical SQIF response under magnetic field $B$. For the sake of clarity, $B$ is the magnetic field after subtraction of a constant ambient field in the un-shielded environment. As shown in Figure \ref{Figure2} \textit{(b)} for a bias current $I=60$ $\mu$A at $T=66$ K, the DC voltage $V$ shows a pronounced anti-peak around zero magnetic field, whose amplitude $\Delta V_{max}=max(V-V_{min})$ (voltage swing) and maximum slope $V_{Bmax}=max\left(\frac{\partial V}{\partial B}\right)$ depend on $I$ and $T$. As already reported\cite{Ouanani:2016cr,Pawlowski:2018jj}, there is an optimal $(I,T)$ couple for which these parameters are maximum, namely $V_{Bmax}\sim 125\ \mathrm{VT}^{-1}$ and $\Delta V_{max}=560\ \mu$V. For this device,  $I_{OPT}=60\ \mu$A and $T_{OPT}=66$ K. Figure \ref{Figure3}\textit{(a)} shows a color-scale plot of the transfer function $V_{B}=\frac{\partial V}{\partial B}$ as a function of $I$ and $B$ at $T_{OPT}$. Two pronounced extrema can be seen, corresponding to optimal field and current conditions to detect a DC magnetic field. In the following, we are studying the ability of such a device to detect the magnetic component of RF waves, and therefore to be used as highly sensitive sensor in the GHz frequency range.

The device is DC biased with $I_{OPT}$ at $T_{OPT}$ and exposed to RF waves at a frequency of $f=1.125$ GHz. The RF power delivered by the source is $P_{RF}=0$ dBm.
As compared to our previous measurements\cite{Pawlowski:2018jj} where the RF wave was directly coupled on-chip, we are in a situation of weak RF coupling. Even at the highest input RF power used here ($P_{RF}=10$ dBm), neither the $I-V$ characteristics nor the $V(B)$ one change with $P_{RF}$ within 1\%.

The output RF voltage of the SQIF is measured with a spectrum analyzer under a swept DC magnetic field $B$. The amplitude of the signal $V_{RF}$ at frequency $f$ is partially modulated by $B$, which is a clear signature of a SQIF response. In Figure \ref{Figure3} \textit{(b)} we plot the pure SQIF response\cite{Pawlowski:2018jj} $\Delta V_{RF}=V_{RF}(B)-V_{RF}(B=0)$ as a function of $B$ (red squares) at $T_{OPT}$. In the same graph is shown the variation of $V_{B}$ (black line), which is a cut of the Figure \ref{Figure3} \textit{(a)} for $I_{OPT}=60\ \mu$A . The two curves superimpose with a very good accuracy as expected in a linear regime\cite{Pawlowski:2018jj}. Indeed, the total magnetic field seen by the SQIF is $B_{TOT}=B+b_{RF}\sin{\left(2\pi f t\right)}$, where $b_{RF}$ is the RF magnetic field amplitude, proportional to $\sqrt{P_{RF}}$. For small $b_{RF}$, one can make a first order Taylor expansion of the output signal, and $V_{RF}\propto \partial V/\partial B=V_{B}$. This is valid for temperatures corresponding to the Josephson regime, as shown in Figure \ref{Figure3} \textit{(d)}. The evolutions of $V_{B}$ and $\Delta V_{RF}$ ($P_{RF}=0$ dBm) with the bias current $I$ also coincide as shown in Figure \ref{Figure3} \textit{(c)} at $T_{OPT}$, in which we have plotted $\Delta V_{RF}^{-}=min(\Delta V_{RF}(B<0))$ and $\Delta V_{RF}^{+}=max(\Delta V_{RF}(B>0))$ as a function of $I$ (red symbols) to account for the relative signs of the magnetic field. On the same graph is shown  (black line) $V_{B}^{-}=min(V_{B}(B<0))$ and $V_{B}^{+}=max(V_{B}(B>0))$ as a function of $I$. This analysis clearly proves that the SQIF response is at play in the RF detection. Indeed, the evolution of the RF signal closely follows that of the DC transfer factor under $B$, $I$ and $T$ changes. This allows us going one step further and making parametric plots of the data to extract more quantitative information.

% Figure 4
\begin{figure*}[ht]
%\begin{minipage}{0.67\textwidth}
%\includegraphics[scale=0.57]{figure4_22012019}%\\
\includegraphics[scale=0.57]{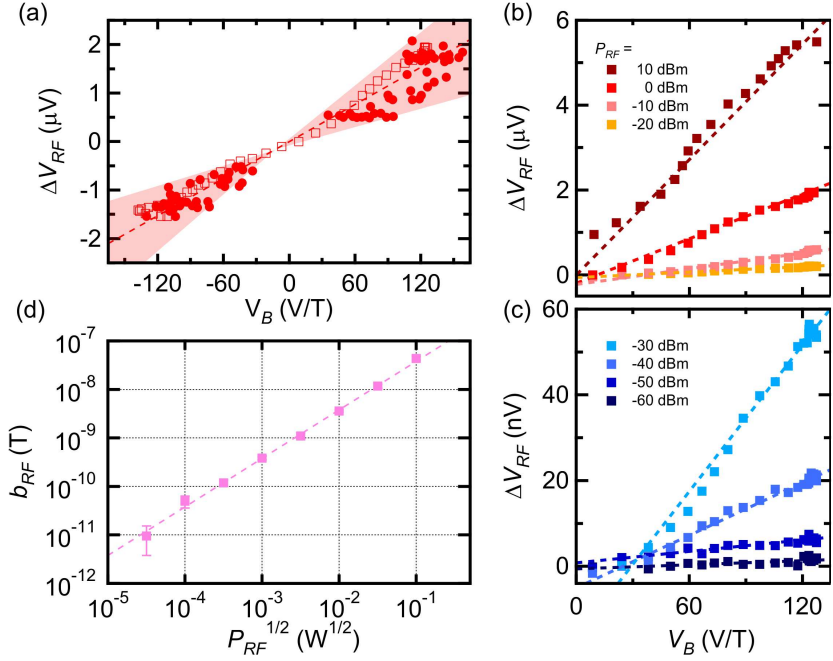}%\\
%\end{minipage}\hfill
 %\begin{minipage}{0.285\textwidth}
  \caption{
  All measurements have been made at $T_{OPT}=66$ K, $I_{OPT}=60\ \mu$A.
\textit{(a)} $\Delta V_{RF}^{-}$ ($P_{RF}=0\ dBm$) as a function of $V_{B}^{-}$, and $\Delta V_{RF}^{+}$ as a function of $V_{B}^{+}$ for different $I$ (solid symbols) and different magnetic fields $B$ (open symbols).  $B$ ranges from $-3.3\ \mu$T to $+2.6\ \mu$T and $I$ from $47\ \mu$A to $91\ \mu$A. The dashed line is the best linear fit of slope $b_{RF}$.
\textit{(b) $\&$ (c)} Same parametric plot in $B$ for different incident RF powers $P_{RF}$. 
\textit{(d)} Detected RF magnetic field $b_{RF}$ as a function $\sqrt{P_{RF}}$. The dashed line is a line of slope 1.
}
  \label{Figure4}
 %  \end{minipage}
\end{figure*}

In the following, we estimate quantitatively the RF magnetic field sensitivity of the device. We express the first order Taylor expansion of the voltage for small $b_{RF}$ as follows $V(B_{TOT})=V(B)+\frac{\partial V}{\partial B} b_{RF} \sin{\left(2\pi f t\right)}+C b_{RF} \cos{\left(2\pi f t\right)}$, where the last term accounts for the regular induction (non-SQIF response) of the device and C is a constant\cite{Pawlowski:2018jj}. The measured $V_{RF}$ at frequency $f$ is the Fourier amplitude of the linear term, and $\Delta V_{RF}=\partial V/\partial B \cdot b_{RF}=V_{B}\cdot b_{RF}$. The amplitude of the RF magnetic field is therefore just the ratio $b_{RF}=\frac{\Delta V_{RF}}{V_{B}}$. In Figure \ref{Figure4} \textit{(a)} $\Delta V_{RF}$ is plotted as a function of $V_{B}$ for different bias currents $I$ (solid symbols) and magnetic fields $B$ (open symbols) at $T_{OPT}$. The RF input power is $P_{RF}=0$ dBm. All the points align on a single straight line (dashed line in the Figure) as expected for this parametric plot. According to the above expression, the slope of this line is the RF magnetic field amplitude, which is here $b_{RF}=12\ \mathrm{nT} \pm 2\ \mathrm{nT}$. The uncertainty is given by the reddish zone in Figure \ref{Figure4} \textit{(a)}. The same parametric plot in $B$ made for different $P_{RF}$ ranging from $-60$ dBm to $10$ dBm also shows a linear behavior (Figure \ref{Figure4} \textit{(b)} \& \textit{(c)} ). The slope of these curves, that is $b_{RF}$, is then plotted as a function of $\sqrt{P_{RF}}$ on a log-log scale in Figure \ref{Figure4} \textit{(d)}. The dashed line has a slope of 1 as expected. This shows that  the RF measured magnetic field is proportional to the input one over 7 decades in RF power, and that the minimum field measured in this series of experiments is of the order of $b_{RF}\sim\ 10\ \mathrm{pT}$. The sensitivity of this SQIF in the $1$ kHz bandwidth of the zero-span mode of our analyzer is therefore $s\sim\ 300\ \mathrm{fT}/\sqrt{\mathrm{Hz}}$.

Such numbers are in line with what is expected. The amplitude of the magnetic field produced in the near-field by an antenna of diameter $d$ on its axis at a distance $D$ is $\mid B \mid=\frac{\mu_{0}d^2I_{RF}}{8D^3}$, where $I_{RF}$ is the RF current in the antenna\cite{Balanis:2005vs}. Taking into account the impedance of the antenna at $f=1.125$ GHz (resistance $= 1.8\ \Omega$, inductance $\times 2\pi f\ =50\ \Omega$) and the reflection coefficient $S_{11}=0.96$ of our set up, we estimate the produced RF magnetic field to be $b_{RF} \sim\ 25\ \mathrm{pT}$ for $P_{RF} = -60$ dBm, which is only twice the measured one. This is the maximum field produced by the emitting antenna, and a more accurate calculation with the exact geometry of the antenna and the SQIF would give a lower value.

The main targeted applications for SQIFs are sensitive detectors of free space RF waves for radars, communications or security systems in open environments, for which the noise floor is typically in the -120 dBm range. The minimum detected signal in conventional devices is therefore around -110 dBm. We can calculate the average power $P$ corresponding to the minimum RF field we detected ($b_{RF}\sim\ 10\ pT$), $P=\frac{b_{RF}^2\epsilon_{0}c^{3}\Sigma}{2}$, where $\epsilon_{0}$ is the air permittivity, $c$ the speed of light and $\Sigma$ the surface of the detector, $\sim\ 1\cdot 10^{-6} m^{2}$ for the present SQIF. We find $P\sim$ -110 dBm, not far from the targeted value, with a device much smaller than regular antennas.

The field sensitivity around $1$ GHz achieved with this device compares favorably with the best ones using atomic magnetometers which operate at room temperature, in the $\mu \mathrm{T}/\sqrt{\mathrm{Hz}}$ range,  reported  by Stark \textit{et al}\cite{Stark:2017jr} and Horsley \textit{et al}\cite{Horsley:2016ce} (up to $50$  GHz), and more recently down to $130 \mathrm{pT}/\sqrt{\mathrm{Hz}}$ at $\sim$ 2 GHz\cite{Andrew:2018fo}. It is worthwhile noticing that this is the highest frequency ever reported for High-$T_c$ SQIF operation.
Moreover, we could observe a SQIF response up to $7.7$ GHz confirming the intrinsic broadband character of the device, but not perform a quantitative analysis of the signal because of poor impedance matching of the whole circuit at this frequency, and the associated numerous parasitic resonances in the detection system. The latter limit the actual bandwidth as well, to roughly 30 MHz at 1.125 GHz.

We can compare these results with the only low-$T_c$ SQUIDs arrays which have been successfully operated as RF antennas in the GHz region so far\cite{Mukhanov:2014ig,Prokopenko:2015dx}. Authors report on "power gains" in the range of 5 to 30 dB at frequencies between $9$ and $12$ GHz mostly, with a bandwidth varying from  $\sim$ 30 to $300\ \mathrm{MHz}$ depending on the configuration. However, what is shown is an ON/OFF operation. Indeed, a comparison is made on the transmission coefficient $S_{21}$ when the SQIF is DC powered (ON), and not powered (OFF). The ratio is measured to be between 5 and 30 dB, and in fact corresponds to the  signal that can be measured, exactly as in our case, and not the true gain of an active system. In addition, no magnetic field dependence is specifically reported, which is the proof of SQIF operation.

Better sensitivity can be achieved with our ion-irradiated HTS SQIF, by increasing the transfer factor $V_{B}$ which is quite low for the device described here ($\sim\ 125$ VT$^{-1}$) as compared to our previous results\cite{Ouanani:2016cr} ($\sim\ 1000$ VT$^{-1}$), or to the best results with step-edge HTS JJ of the CSIRO group $V_{B}\sim\ 1750$ VT$^{-1}$\cite{Mitchell:2016in} and $V_{B}\sim\ 40$  kVT$^{-1}$\cite{Keenan:2017} using more complex architectures. A new design using 2D arrays is under test for enhanced transfer factor. In addition, one can improve the sensitivity by concentrating the magnetic field. The actual flux focusing factor of the individual SQUIDs is of the order of 3 in the actual geometry\cite{Ouanani:2015tw}, which can be slightly increased. We can also put large superconducting pads in the vicinity of the SQIF to concentrate the flux even further\cite{Labbe:2018iy}.

In summary, we have studied the RF properties of a HTS 1D series SQIF array made of ion irradiated JJ, and tested its performance as a sensitive magnetometer in the GHz frequency range, in an un-shielded magnetic environment. Operating in the $60-68$ K temperature range, the device showed a SQIF response under DC magnetic field, and could detect RF electromagnetic waves emitted by a loop antenna up to $7.7$ GHz. We evidenced that the applied DC magnetic field modulates the RF output signal, sign of a SQIF operation, and that the absolute value of the RF magnetic field can be extracted from the measurement at $1.125$ GHz. At optimum conditions, we have shown that the device can detect an RF magnetic field of about $10\ \mathrm{pT}$ at this frequency, corresponding to a sensitivity of $\sim\ 300\ \mathrm{fT}/\sqrt{\mathrm{Hz}}$, in a bandwidth of 30 MHz and with a dynamical range of 70 dB. Such result paves the way of highly sensitive RF magnetometers working at temperatures where cost- and energy-effective cryo-coolers operate, which are broadband in frequency and sub-wavelength in size. These are three major issues for a wide range of applications where compact RF antennas are required.\medskip 

The authors thank Yann Legall (ICUBE laboratory, Strasbourg) for ion irradiations, St\'{e}phane Hol\'{e}, Thierry Ditchi, Emmanuel G\'{e}ron and J\'{e}r\^{o}me Lucas for fruitful discussions and technical help. This work has been supported by ANRT and Thales through a CIFRE PhD fellowship (2015/1076), the QUANTUMET ANR PRCI program (ANR-16-CE24-0028-01), the T-SUN ANR ASTRID program (ANR-13-ASTR-0025-01), the Emergence Program from Ville de Paris and by the R\'{e}gion Ile-de-France in the framework of the DIM Nano-K and Sesame programs. 

%\nocite{*}
\bibliography{SQIF_RF_Magnetometer}% Produces the bibliography via BibTeX.

\end{document}